\documentclass[twocolumn,showpacs,superscriptaddress, preprintnumbers , amsmath,amssymb,letterpaper]{revtex4}
\usepackage{dcolumn}% Align table columns on decimal point
\usepackage{bm}% bold math
\usepackage{amsmath}
\usepackage{multirow}
\usepackage{graphicx}
\begin{document}

 %\twocolumn
 %\baselineskip = 1.5\baselineskip

 \newcommand{\re}{\mathop{\mathrm{Re}}}
 \newcommand{\im}{\mathop{\mathrm{Im}}}
 \newcommand{\D}{\mathop{\mathrm{d}}}
 \newcommand{\I}{\mathop{\mathrm{i}}}
 \newcommand{\E}{\mathop{\mathrm{e}}}
 \newcommand{\unite}[2]{\mbox{$#1\,{\rm #2}$}}
 \newcommand{\myvec}[1]{\mbox{$\overrightarrow{#1}$}}
 \newcommand{\mynor}[1]{\mbox{$\widehat{#1}$}}
 \newcommand{\rmsemit}{\mbox{$\tilde{\varepsilon}$}}
 \newcommand{\mean}[1]{\mbox{$\langle{#1}\rangle$}}

\title{Numerical modeling of a table-top tunable Smith-Purcell  \\
Terahertz  free-electron laser operating in the superradiant regime }
% Force line breaks with \\
\author{C. Prokop} \affiliation{Northern Illinois Center for
Accelerator \& Detector Development and Department of Physics,
Northern Illinois University, DeKalb IL 60115,
USA}
\author{P. Piot} \affiliation{Northern Illinois Center for
Accelerator \& Detector Development and Department of Physics,
Northern Illinois University, DeKalb IL 60115,
USA}\affiliation{Accelerator Physics Center, Fermi National
Accelerator Laboratory, Batavia, IL 60510, USA}
\author{M. C. Lin}\affiliation{Tech-X Corporation, Boulder, CO 80303, USA}
\author{P. Stoltz} \affiliation{Tech-X Corporation, Boulder, CO 80303, USA}
\preprint{PREPRINT-FERMILAB-PUB-10-033-APC}
\date{\today}% It is always \today, today,

\begin{abstract}
Terahertz (THz) radiation occupies a very large portion of the electromagnetic spectrum and has generated much recent interest due to its ability to penetrate deep into many organic materials without the damage associated with ionizing radiation such as x-rays. One path for generating copious amount of tunable narrow-band THz radiation is based on the Smith-Purcell free-electron laser (SPFEL) effect. In this Letter we propose a simple concept for a compact two-stage tunable SPFEL operating in the superradiant regime capable of radiating at the grating's fundamental bunching frequency.  We demonstrate its capabilities and performances via computer simulation using the conformal finite-difference time-domain electromagnetic solver {\sc vorpal}.
\end{abstract}
                              %display desired%
\maketitle
Terahertz (THz) radiation is finding use in an increasingly wide variety of applications including medical imaging, homeland security and global environment monitoring~\cite{masayoshi}. Increasing access to THz technologies requires the development of compact and tunable THz sources.  Recent years have witnessed a resurgence of interest in Smith-Purcell free-electron lasers (SPFELs) operating as a backward wave oscillator~\cite{Brau1, Kumar1} following on an idea initially discussed in Ref.~\cite{Wachtel}. The developed model was recently benchmarked by laboratory experiments~\cite{Brau2009, Donohue2009}. THz sources based on SPFELs are foreseen to have table-top footprint and can operate in a continuous wave mode, enabling the production of moderate average output power (on the order of Watts).  

In an SPFEL, a low energy ($\sim 50$~keV) sheet DC electron beam is propagated close to a metallic grating with velocity $\mathbf{v}\equiv c \beta \hat y$ . The beam excites  evanescent surface waves with axial field of the form $E_{y,e}(x,y)=E_{0,e} \exp(\alpha x)$ where $\alpha \equiv 2\pi /(\beta\gamma \lambda_{e})$ and $\gamma\equiv (1-\beta^2)^{-1/2}$ is the Lorentz factor. The evanescent wave can, under certain circumstances, have a negative group velocity~\cite{Brau1}. In such a case the wave counter-streams the electron beam direction and can couple to the beam, thereby giving rise to an energy modulation. Due to the non-relativistic nature of the beam ($\gamma\simeq 1$), the impressed  energy modulation eventually transforms into a density modulation at wavelength $\lambda_{e}$. The produced microbunches will result in strongly enhanced radiation at harmonic frequencies of the microbunching frequency $f_{e}\equiv c/\lambda_{e}$.  In an SPFEL the radiative mechanism is the Smith-Purcell (SP) effect~\cite{SmithPurcellOriginal}. If instead of a DC electron beam, a beam consisting of a train of microbunches is used the SP radiation is emitted in the super-radiant regime~\cite{Brau2, dli,kesar} in which the radiation rate goes as th number of electrons in each microbunch squared. Prebunching the electron beam in a way that satisfies emission of super-radiant radiation is however challenging and a possible solution discussed in, e.g., Ref.~\cite{yli} significantly decreases the average power capability of the SPFEL. 

\begin{figure}[hhhhh!!!!!!!!!!!!]
\centering
\includegraphics[width=0.45\textwidth]{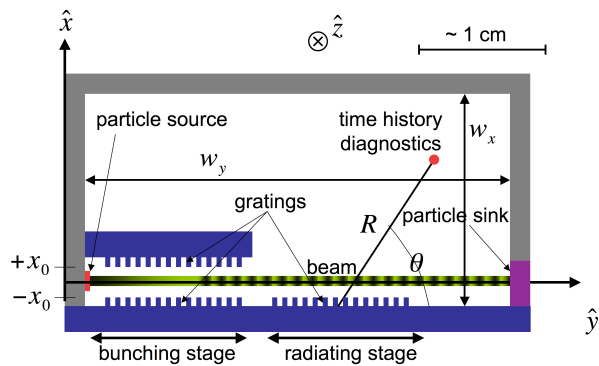}
\caption{(Color online) Diagram of the two-stage SPFEL. The rectangular box represents the 3D computational domain used in the {\sc Vorpal} simulations. Grey and blue blocks respectively stand for perfectly matched layer and metallic boundaries. }
\label{fig:Triple_Geometry}
\end{figure}

In this Letter we consider and present detailed numerical simulations of a ``two-stage" SPFEL; see Fig.~\ref{fig:Triple_Geometry}. In such a configuration a first stage, referred to as a ``buncher'', is optimized to enhance the beam-evanescent wave interaction, thereby resulting in faster bunching than in the configuration analyzed in previous papers~\cite{Brau1,Kumar1,shi}. A second stage, referred to as a ``radiator", consists of a grating with parameters tuned to produce coherent SP radiation at frequencies $nf_{e}$ ($n$ is an integer). 
 
The numerical simulations were performed using {\sc vorpal}, a conformal finite-difference time-domain (CFDTD) particle-in-cell electromagnetic solver~\cite{vorpal}. {\sc vorpal} is a parallel, object-oriented framework for three dimensional relativistic electrostatic and electromagnetic plasma simulation. We extensively benchmarked our initial {\sc vorpal} simulations of the SPFEL process against earlier work~\cite{Donohue1}; see Ref.~\cite{ProkopPiotFEL2009}. The geometric configuration of our simulation model is shown in Fig.~\ref{fig:Triple_Geometry}. The model includes perfectly-conducting rectangular gratings, a particle source on the left end side and a ``particle sink" on the right that allows macroparticles to exit the computational domain without being scattered or creating other source of radiation. The grating grooves are along the $\hat{z}$ direction and the beam propagates along the $\hat{y}$ axis. Beside the grating, the other surrounding boundaries consist of perfectly matched layers that significantly suppresses artificial reflections of incident radiation.  The particle source produces a uniformly-distributed (in all directions) DC beam with an instantaneous risetime and is transversely confined by an uniform external axial magnetic field $\mathbf{B}_a=B_a \hat{y}$. The parameters used for the simulations presented below are gathered in Tab.~\ref{tab:Grating_Parameters}, with $E= 50$~keV, $I=135$~A.m$^{-1}$, $b=50$~$\mu$m, and $g=20$~$\mu$m; for the sake of simplicity the buncher and radiator gratings are identical. 

\begin{table}
\caption{\label{tab:Grating_Parameters} Grating and beam parameters used for the {\sc Vorpal} simulations.}
\begin{center}
\begin{tabular}{l  c c  c}
\hline \hline
Parameter & Symbol & Value & Unit \\ 
\hline
Grating Period & $\lambda_g $ & $200$ & $\mu$m \\
Groove Width & $w$ &$100$ & $\mu$m \\
Groove Depth & $h$ & $100$ & $\mu$m \\
Number of Periods/grating & $N_g$ & $75$ &  $--$\\
Electron Energy & $E$ & $25$-$200$  & keV \\
Beam Current & $I$ & $\geq$~$135$ & A/m \\
Beam Thickness & $b$&  $20$-$400$  &  $\mu$m \\
Beam Clearance & $g$ & $20$-$30$ & $\mu$m \\
Grating gap & $2 x_0$ & $b+2g$ & $\mu$m \\
External Magnetic Field & $B_a$ & $2.0$ & T \\

\hline \hline
\end{tabular}
\end{center}
\end{table}

\begin{figure}[hhhhh!!!!!!!!!!!!]
\centering
\includegraphics[width=0.47\textwidth]{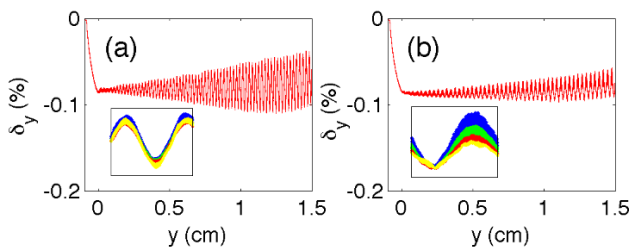} %32, 33
\caption{(Color online) Snapshot of the axial phase spaces $(y, \delta_y)$ (where $\delta_y= \gamma\beta_y/[\beta_y (y=0)\gamma (y=0)]-1$ is the fractional axial momentum spread) for a double (a) and single (b) grating taken at 151~ps. The lower insets on each of the plot present zoomed-in phase space over for $y\in [0.50, 0.53]$~cm, for which the macroparticles have been color-coded accordingly to their height within the beam. The double-grating simulation has less distinct layers, signifying more uniform bunching across the height of the electron beam. The beam parameters are  $E= 50$~keV and $I=135$~A.m$^{-1}$.\label{fig:Phase_Number_Comparison}}
\end{figure}
%Runs 32, 33, dump1

\begin{figure}[hhhhh!!!!!!!!!!!!]
\centering
\includegraphics[width=0.45\textwidth]{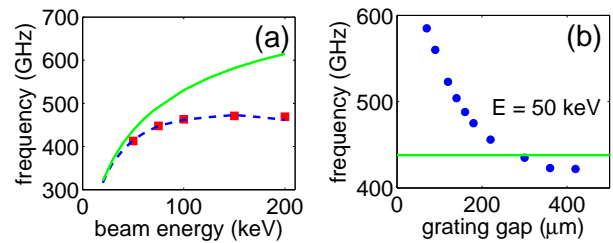}
\caption{(Color online) Evanescent wave frequency versus beam energy for the single-grating configuration (a) and evanescent wave frequency versus grating separation for a double-grating configuration (b). The markers are results from {\sc vorpal} simulations while the dashed line is obtained from the dispersion relation derived in Ref.~\cite{Brau1}.  In both graphs, the green solid line denotes the minimum allowed SP frequency for a grating period of 200~$\mu m$.}\label{fig:Phase_Energy_Comparison}
\end{figure}

We first analyze the buncher section of the proposed two-stage SPFEL. Contrary to the SPFEL configurations explored in previous work, the beam propagates between two identical gratings arranged symmetrically; see Fig.~\ref{fig:Triple_Geometry}.  Considering the two-dimensional problem, the evanescent waves produced from the beam passing over each grating located at $\pm x_0$ are $E_{y,e}^{\pm}(x, y) = {E_{0,e}}(y)e^{-\alpha(x_0\mp x)}$ (where $\pm$ respectively designate the electric field contributions from the upper and lower grating).  The overlap of the evanescent waves from both gratings results in an evanescent field $ E_{y,e} (x,y) = 2{E_{0,e}}(y) e^{-\alpha x_0}\cosh(\alpha x)$. Beside being stronger by a factor of two (for $x=0$) thereby bunching the beam stronger,  the field is also more uniform across the beam's thickness and effectively result in a more uniform bunching. The strengthened evanescent field may also be taken advantage of to relax the start current requirements. These features are confirmed by numerical simulation; see Fig.~\ref{fig:Phase_Number_Comparison}.  The velocity modulation is approximately twice as large for the double grating configuration than for a single grating configuration. The presence of the second grating also affects the dispersion relation and $\lambda_e$, which in-turn leads to a different bunching frequency than the single grating system; see analysis in Ref.~\cite{Freund}.  This feature offers a greater flexibility for tuning the evanescent wave frequency (and therefore the radiation frequency) by either varying the electron beam energy, or altering the gap between the two gratings as demonstrated in Fig.~\ref{fig:Phase_Energy_Comparison}. As the gap increases, the frequency converges with that of a single-grating sytem. In addition the double grating configuration allows the emission of SP radiation at the bunching frequency. Over the considered range of energy $E\in[25, 200]$~keV and grating gap $2 x_0 \in [90,440]$, significant velocity modulation is observed and the corresponding evanescent wavelength variation of respectively $\sim 15$~\% and $\sim 35$~\% are approximately two orders of magnitude than the produced radiation spectrum width as will be shown later.

The second stage of the proposed SPFEL consists of passing the microbunched beam close to a single grating. The spectral intensity radiated by a bunch of $N$ electrons via the SP effect is related to the single electron intensity  $\frac{dI}{d\omega}\big|_1$ via $\frac{dI}{d\omega }\big|_N=\frac{dI}{d\omega }\big|_1 [N+N^2|S(\omega)|^2]$ where $S(\omega)$ is the intensity-normalized Fourier transform of the normalized charge distribution $S(t)$~\cite{saxon}. Considering a series of $N_b$ identical microbunches  with normalized distribution $\Lambda(t)$  we have $S(t)=N_b^{-1} \sum_{n=1}^{N_b} \Lambda(t+nT)$ (where $T=\lambda_e/ c$ is the period) giving  $|S(\omega)|^2 =  \xi  |\Lambda(\omega)|^2$. The intra-bunch coherence factor  $\xi\equiv N_b^{-2} \sin^2(\omega N_b T/2)/[\sin^2(\omega T/2)]$ describes the enhancement of radiation emission at resonance, i.e. for frequencies $\omega = 2\pi n c/\lambda_e$.

\begin{figure}[hhhhh!!!!!!!!!!!!]
\centering
\includegraphics[width=0.44\textwidth]{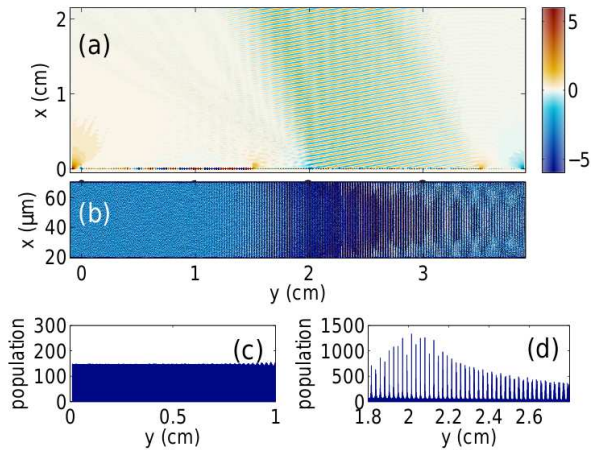}
\caption{(Color online) Snapshots at $t=906$~ps of the axial electric field $E_y$ and corresponding spatial distribution of the beam in the $(x,y,z=0)$ plane for a two-stage SPFEL (the colorbar units are MV.cm$^{-1}$). The associated axial projections at two axial locations are shown in  (c) and (d). }\label{fig:Triple_Grating_Ey}
\end{figure}

Figure~\ref{fig:Triple_Grating_Ey} shows a contour plot of the $E_y$-field over the computational domain and clearly demonstrates the emission of SP radiation with a planar wavefront from the second grating. The planar wavefront is a signature of superradiance due to constructive interference between successive bunches.  The wavevector associated to this radiation makes a $106^{\circ}$ angle and the wavelength value $\simeq 540$~$\mu$m is consistent with the one exected from the SP relation~\cite{SmithPurcellOriginal}  $\lambda_m=\lambda_g/|m| (\beta^{-1}-\cos\theta)=540~\mu$m  (taking $m=1$, $\lambda_g=200$~$\mu$m and $\beta=0.4127$).  Whereas the single-grating SPFEL requires that the super-radiant emission occurs at $m\ge  2$, the proposed two-stage SPFEL can operate in the super-radiant mode at $m=1$ by a proper selection of parameters.  Even if the radiator and the buncher have the same grating period, the increased bunching frequency from the double-grating dispersion relation may match that of the m=1 mode of the SP relation.  Operating at $m=1$ enables the generation of higher radiation rates compared to $m \ge 2$.  Figure~\ref{fig:Triple_Grating_Ey} also displays a snapshot of the spatial distribution and associated projections confirming the formation of microbunches. 

\begin{figure}[hhhhh!!!!!!!!]
\centering
\includegraphics[width=0.46\textwidth]{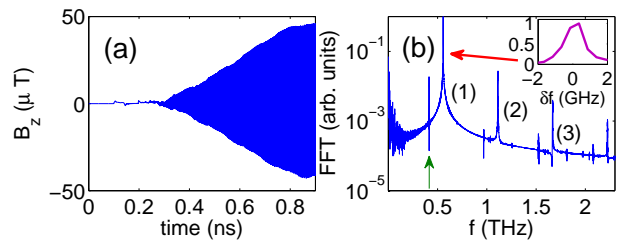}
\caption{(Color online) Time evolution of the magnetic field $B_z$ at ($R,\theta$)=(2~cm, 106$^{\circ}$) (a) and associated FFT (b). The beam parameters are  $I=135$~A.m$^{-1}$ and $E=50$ keV. The inset in (b) is a zoom-in of the 1st harmonic peak showing the bandwidth of 1.3~GHz (FWHM). The numbers (1), (2) and (3) indicate the harmonics of the evanescent wave associated to the double grating configuration. The green vertical arrow indicates the evanescent wave supported by the downstream grating, which is below the minimum frequency allowed by SP radiation and thus only scatters off the edges of the grating.}\label{fig:History_Comparison}
\end{figure}

The time evolution of the magnetic field $B_z$ recorded at $\theta=106^{\circ}$ and $R=2$~cm along with its spectrum are presented in Fig.~\ref{fig:History_Comparison}. The spectrum shows that significant emission occurs at $f_e=556$~GHz, the frequency of the evanescent wave supported by the buncher section, and its harmonics. The bandwidth of the radiation emitted at the first harmonic is $\delta f \simeq 1.3$~GHz resulting in $df/f_e \simeq 0.23$~\%.  The spectrum also displays a  413~GHz frequency component which corresponds to the evanescent wave produced by the downstream grating. Taking the  wavefront to be a plane wave with steady-state peak magnetic field of  $B_0 \sim 40$~$\mu$T (see Fig.~\ref{fig:History_Comparison}), we estimate the time-averaged Poynting vector to be $\langle S \rangle = cB_0^2/(2\mu_0) \simeq 2 \times 10^4$~W.m$^{-2}$. Harvesting the radiation using a  $1\times 1$~cm$^2$ mirror would result in a total average power of $P\sim 2$~W.  

In summary we have demonstrated several advantages of a two-stage SPFEL and show that the device can be used to produce tunable super-radiant THz radiation with average power of the order of Watts. 

We thank Dr. P. Spentzouris (FNAL) for granting us access to the NERSC computers where our simulations were performed. This work was partially supported by the US Department of Education under contract P116Z050086 with Northern Illinois University.

\end{document}